\documentclass[fleqn,usenatbib]{rasti}

\usepackage{newtxtext,newtxmath}
\usepackage{fix-cm}

\usepackage[T1]{fontenc}

\DeclareRobustCommand{\VAN}[3]{#2}
\let\VANthebibliography\thebibliography
\def\thebibliography{\DeclareRobustCommand{\VAN}[3]{##3}\VANthebibliography}

\usepackage{graphicx}	
\usepackage{amsmath}	

\title[Enhancing Fast Radio Transient Detection]{Enhancing Fast Radio Transient Detection with Mask R-CNN Image Segmentation}

\author[S. Belmonte Diaz et al.]{
Sergio Belmonte Díaz,$^{1}$\thanks{E-mail: sergio.belmontediaz@manchester.ac.uk (SBD)}, Rene P. Breton$^{1}$, Zafiirah Hosenie$^{2}$, Ben W. Stappers$^{1}$
\\
$^{1}$Jodrell Bank Centre for Astrophysics, Department of Physics and Astronomy, The University of Manchester, Manchester M13 9PL, UK\\
$^{2}$Microsoft, 21 Station Road Cambridge, CB1 2FB, United Kingdom\\
}

\date{Accepted XXX. Received YYY; in original form ZZZ}

\pubyear{2025}

\begin{document}
\label{firstpage}
\pagerange{\pageref{firstpage}--\pageref{lastpage}}
\maketitle

\begin{abstract}
Traditionally, fast radio transient searches are conducted on dedispersed time series using thresholding techniques based on the statistical properties of the data. However, peaks in dedispersed time series do not directly provide information on the nature of the source. In the DM–time domain, the S/N variation of real, dispersed astrophysical signals forms a characteristic bow tie shape, whereas radio frequency interference (RFI) can take multiple different forms. We have developed a method that bypasses the thresholding step of traditional single-pulse searches in favour of a direct DM–time domain image analysis. The backbone of our pipeline is a Mask R-CNN, a deep learning model designed for object detection, enabling it to identify the bow tie signature and distinguish real sources from RFI. Previous deep learning models often include a snippet of the DM–time domain in their input. We have trained the model on simulated bursts injected on top of real MeerKAT noise observations. We tested the model on MeerKAT follow-up observations of the repeater FRB20240114A and we were able to recover all bursts with a signal-to-noise above the traditional threshold, while detecting two bursts that were fainter. Our new approach considerably reduces the number of candidates above a nominal threshold while being capable of running in real time for typical surveys. We also propose a modified version of the traditional dedispersion plan optimised for this method.
\end{abstract}

\begin{keywords}
Machine Learning -- Software -- Algorithms
\end{keywords}



\section{Introduction}

\begin{figure*}
	\includegraphics[width=\textwidth]{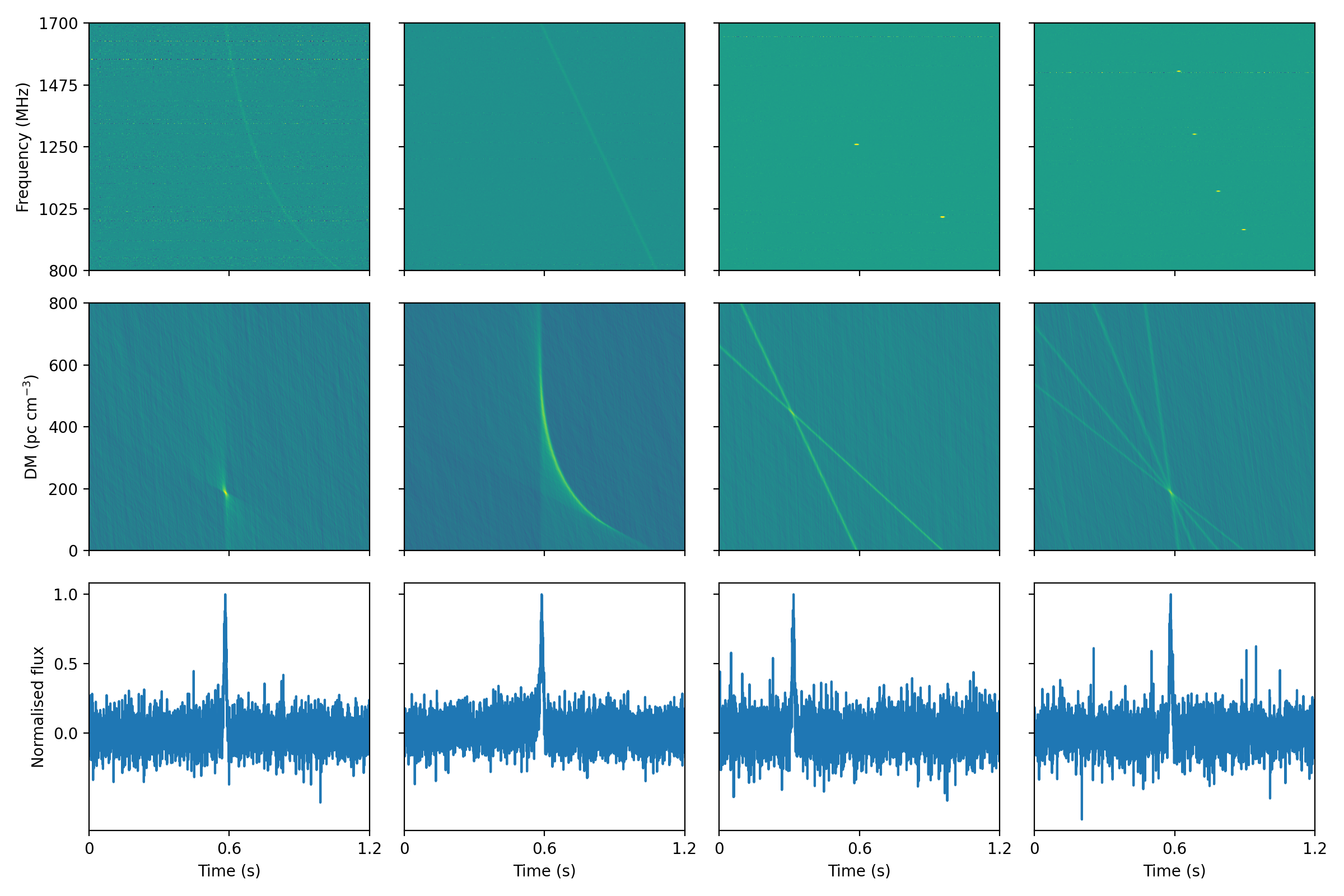}
    \caption{The top row displays frequency-time plots for different sources, the middle row shows the DM-time transform of these sources, and the bottom row presents the frequency-summed time series at the point of maximum S/N. The first column depicts a broadband burst dispersed according to Equation \ref{eq:time-dispersion}, simulating an ideal transient source. The second column illustrates a linear frequency-drifting source, representing a type of RFI. The third column shows narrowband RFI at two random frequency channels, a common occurrence in radio transient searches. The fourth column represents a narrowband, intermitent burst following the dispersive relation expected from astrophysical sources, illustrating a more complex emission scenario than the ideal case in the first column.}
    \label{fig:rfi_in_DM_space}
\end{figure*}

The search for short-timescale radio transients has led to the discovery of exotic astrophysical phenomena such as Rotating Radio Transients (RRATs) \citep{2006Natur.439..817M} and Fast Radio Bursts (FRBs) \citep{2007Sci...318..777L}. RRATs are pulsars that, due to the extremely variable nature of their pulse amplitudes, are easier to detect through a single pulse search rather than the typical periodicity searches \citep{2011MNRAS.415.3065K}. These are examples of extreme nulling pulsars, which only emit detectable bursts once every hundreds or thousands of rotations \citep{2011BASI...39..333K}. There is significant interest in searching for RRATs since their erratic emission behaviour -- which deviates from the more regular nature of most pulsars -- may provide insights about processes powering their radio emission. Incidentally, single pulse search techniques have led to the discovery of a pulsar with a spin period of 76 seconds, which challenges the traditional understanding of emission mechanisms and neutron star spin evolution \citep{2022NatAs...6..828C}.

Fast Radio Bursts are extremely bright ($\sim$ 1 Jy ms), millisecond-duration ($\sim$ 1 ms) radio pulses of extragalactic origin ($z \sim 0.5$) \citep{2021ApJS..257...59C}. Although hundreds of FRBs have been detected \citep[see, e.g.,][]{2013Sci...341...53T, 2018MNRAS.475.1427B,2018Natur.562..386S,2021ApJS..257...59C}, their progenitor and emission mechanism are not yet understood. More than 60 FRBs are reported to be repeaters, rather than single events, in the CHIME repeating FRB database\footnote{\url{https://www.chime-frb.ca/repeaters}} \citep{CHIME2019ApJ,FONSECA2020,2023ApJ...947...83C}, including two that exhibit periodic activity \citep{FRB180916.J0158+65_periodicity,Rajwade2020_FRB121102_periodicity} and a third source showing periodicity at the 3.5$\sigma$ level \citep{Zhou2025}. These observations rule out cataclysmic events as the underlying root cause for at least this subset of the population. However it is not yet clear whether all FRBs repeat \citep{2019MNRAS.484.5500C}, and which fraction of repeaters show periodic activity. FRBs show a diversity of burst properties: single narrow or single broadband components, multiple components detected at the same frequency range or finite-bandwidth component frequency drifting (aka the `sad trombone' effect), and nanosecond microstructure components \citep{Hessels2019,2021ApJ...923....1P,2021ApJ...919L...6M}.

It is clear that finding more FRBs would help deepen our understanding of their progenitors and the physics behind their emission mechanisms, while also enabling population studies aiming to characterise these sources. More importantly, it keeps the door open to discovering new classes of transients related to other astrophysical phenomena. In this regard, efforts have been made towards improving surveying speed by increasing either one or a combination of field of view, sensitivity, and on-sky time. For example, significant progress has been accomplished through the commissioning of commensal transient detection systems capable of operating without disrupting regular science operations \citep[see, e.g.,][]{2010Macquart_CRAFT_ASKAP,2018Law_RealfastVLA,2018MeerTRAP,2018CHIME_Systemreview,DSA2024}. Developments on the survey front, however, come at a cost: more data need to be analysed, and their outputs produce more candidate transients that require veting. Adding to this big data problem is the desire to identify transients in (quasi-)real time so that low-level data products from the telescope can be saved for enhanced post-facto analysis and follow-up observations be triggered if needed. As a result, significant emphasis has gradually been placed on improving the detection pipelines used in radio transient searches so they can run faster and produce outputs of better quality (i.e. be more resilient to false-positives, and recover a more complete sample).

In this paper, we propose a new end-to-end deep learning pipeline for the detection of short-timescale radio transients, where the search is conducted directly in the dispersion measure (DM) and time plane. This bypasses the traditional approach consisting of a match-filtering and thresholding, followed by a machine learning sifting. By utilising the full available information, this approach aims to reduce the number of false candidates and possibly push down the sensitivity limit for detection. The pipeline performs an untargeted search for transients in which a radio observation is dedispersed, sliced into individual images, and searched for transients using a Mask R-CNN model \citep{2021PhDTZafiirah}. The paper is organised as follows: In \S\ref{sec:CurrentMethod}, we present a brief overview of the current search methodology as to provide the necessary background to understand the key features introduced in our approach. Then, in \S\ref{sec:Slicing} we introduce the challenges and proposed solutions to search for radio transients in the DM-time space. We present in \S\ref{sec:DeepLearning} our Mask R-CNN deep learning model and the data generation to train the neural network. Finally, in \S\ref{section:Results}, we report the results of the training and discuss our benchmarking analysis of the execution time of our method, together with its performance at real-time detection of transients in observations.

\section{Overview of the current searching methodology}
\label{sec:CurrentMethod}
Single-pulse searching was described as a match-filtering exercise by \citet{2003ApJ...596.1142C}. Current pipelines are based on algorithms that use thresholding techniques to obtain a list of candidates with a signal-to-noise ratio (S/N) above a certain level of significance (e.g. {\sc presto} \citep{2011ascl.soft07017R}, {\sc heimdall }\citep{2024ascl.soft07016B}, {\sc astroaccelerate} \citep{2020ApJS..247...56A}). Before searching for any potential astrophysical signal, the spectral data is corrected for the frequency-dependent delay induced by the cold ionised interstellar medium (ISM). The ISM is a dispersive medium, which causes a frequency dependency on the group velocity of the electromagnetic waves traveling through it \citep{2012hpa..book.....L}. The effect is a smearing of the signal across time, which reduces the significance of the pulse if not corrected for. The dispersive delay $t(\nu)$ can be characterised by 
\begin{equation}
\label{eq:time-dispersion}
    t(\nu) = \mathcal{D} \times \frac{\textrm{DM}}{v^{2}} \,,
\end{equation} where $\nu$ is the radio frequency, DM is the dispersion measure and $\mathcal{D}$ is the dispersion constant. The DM measures the integrated  content of free electrons along the line of sight between the observer and the source, and is defined as 
\begin{equation}
    \textrm{DM} = \int n_{\textrm{e}} {\rm d}l \,,
\end{equation} where $n_{\textrm{e}}$ is the electron density. The dispersion constant $\mathcal{D}$ is equal to
\begin{equation}
    \mathcal{D} = \frac{e^{2}}{2\pi mc} = 4.1488 \times \textrm{10$^{3}$ MHz$^{2}$ pc$^{-1}$ cm$^{3}$ s} \,,
\end{equation} where $m$ and $e$ are the electron mass and charge respectively, and $c$ is the speed of light.

When searching for new transients, the DM is not known a priori, meaning that many DM trial values need to be searched. In a generic pipeline, for each DM trial, a time series is generated by summing the signals across all frequency channels. Slow fluctuations in the mean level of the time series, due to radio frequency interference (RFI) or instrumentation effects, can hinder the detection of pulses above the noise level. Typically, the baseline is subtracted from the time series before searching for single pulses by applying a robust clipped mean, in which a running mean is used to estimate the baseline, outliers are removed, and the process is iterated \citep{2012PhDT.......465B}. A robust estimate of the root mean square (RMS) of the noise is needed to perform an accurate thresholding. The median absolute deviation (MAD) can be used as a statistic that mitigates the effect of outliers in the time series. The RMS can then be approximated by \mbox{RMS = 1.4862 MAD}, where the constant arises from the fact that the MAD represents the 75th quantile of a normal distribution \citep{2012PhDT.......465B}. After the baseline-removal and the normalisation, the pipeline proceeds by applying a matched filter to the dedispersed time series, which involves convolving the data with a kernel to maximise the S/N of the underlying signal. The kernel is a template of the shape of the signal to find. Single bursts from radio transients show a diversity in pulse shape, but they can be approximated by boxcar filters. The signal-to-noise ratio increases as the filter width approaches the width of the underlying signal. This adds another dimension to the parameter space to search, as the width of the burst is not known a priori. Candidates with a S/N greater than a certain threshold are then selected for further scrutiny.

With the advances in telescope technology, such as MeerKAT \citep{2016mks..confE...1J}, CHIME \citep{2018ApJ...863...48C}, and the future SKA \citep{2015aska.confE..51F}, the volume of data produced has increased to levels where complete storage of dynamic spectra for offline searches becomes impractical. Only candidates selected through real-time software are saved for further inspection. An increase in the search parameters space, notably involving additional beams, dispersion measures, and widths, makes manual inspection of these candidates unfeasible. For this reason machine learning (ML) techniques have been developed to help reduce the number of candidates to be manually inspected by using key features from the candidates to classify real sources and radio frequency interference (e.g. \citet{2016PASP..128h4503W,Farah2019,2019Natur.566..230C,Wu2019,2023MNRAS.518.1629L}).

However, the pipelines to search for single pulses described above do not use all the available information encapsulated in the signal. Applying a threshold on the time series captures all bursts above a certain level of significance, but it does not retain information about their frequency structure, which greatly increases the number of false positives in the search. A key factor in differentiating real astrophysical sources from RFI is the S/N degradation of the signal as a function of time and DM. A broadband, dispersed astrophysical signal suffers a S/N degradation presenting a characteristic bow tie shape in the DM-time space. Conversely, random noise fluctuations, frequency drifting or narrow-band RFI are generally characterised by a different shape in the DM-time plane, as can be seen in Figure \ref{fig:rfi_in_DM_space}. In DM-time space, all four signals shown in Figure \ref{fig:rfi_in_DM_space} appear with distinguishable features, whereas their signature in the summed time series is almost identical. Projecting the data into the DM-time space therefore preserves more information about the nature of the candidate than the dedispersed time series, which could be used to improve the detection of real sources and the rejection of false candidates. It also offers an advantage in distinguishing real sources from zero-DM RFI since these signals appear in the lowest part of the DM-time plane and can be isolated, thus reducing their impact on the rest of the observation.

Even though DM-time domain data provides more information, differentiating between real sources and noise can still be challenging. Narrowband RFI, provided it has enough power, can resemble the S/N degradation of a narrowband signal from astrophysical origin, as seen in the last two columns of Figure \ref{fig:rfi_in_DM_space}. More modern approaches use Deep Learning techniques to assess the nature of the candidates using the DM-time plane as an input. \citet{Connor2018} built a hierarchical hybrid classifier that independently extracts features from up to four input data products: dedispersed dynamic spectrum, pulse profile, DM-time transform, and multibeam detections. The extracted features are then combined into a unified classifier network. Fast Extragalactic Transient Candidate Hunter (FETCH, see \citep{2020MNRAS.497.1661A}) uses two independent Convolutional Neural Networks (CNN) to process a snapshot of the frequency-time and DM-time information of the candidate. However, these ML techniques are only applied after the thresholding search of the dedispersed time series has been done, to further scrutinise the candidates, meaning that most of the search is not done with all the available information.

\section{Searching for candidates in the DM-time plane}
\label{sec:Slicing}

The bow tie structure exhibited by short-timescale astrophysical radio emission in the DM-time plane can vary depending on the burst morphology and, most importantly, the data processing strategy used to obtain the DM-time transform. Three main attributes are relevant: the size of the S/N degradation, its gradient evolution, and its aspect ratio. Below is a short qualitative description of how they are affected.
\begin{itemize}
    \item When the signal is dedispersed at the correct value of DM, its power is coherently summed across multiple frequency channels, maximising the S/N. The real DM and time of the burst marks the centre of the bow tie, while its spread in time and frequency is commensurate to the width of the original burst. At slightly incorrect values of DM, higher frequencies are less affected than lower frequencies by the dispersive delay, meaning that part of their power is still coherently summed. For this reason, under-dedispersed signals will peak later compared to the expected value, while over-dedispersed signals will peak earlier, thus producing the characteristic bow tie shape. Finally, the inherent S/N of the burst also affects the apparent extent of the bow tie, since a higher S/N one would be seen at larger incorrect DM values before fading away.

    \item The exact behaviour of the S/N degradation as a function of time and DM depends on the spectrum and morphology of the burst. A steeper spectrum, for instance, would present more power in the bow tie at later times rather than at earlier times since there is more power in the lower frequencies. Moreover, if the emission is narrow-band, S/N is only present at a small range of frequency channels, which results in the bow tie increasingly resembling a narrow diagonal line. Other effects, such as scintillation or scattering can further distort the shape of the bow tie.

    \item The DM and time range selected to produce the DM-time transform affect the aspect ratio of the signal. If many DM bins are sampled, the S/N degradation will appear to be compressed, and vice versa. Moreover, the reference frequency used to perform the dedispersion will change the orientation of the bow tie. In our work, we have used the frequency channel at the top of the band as our reference frequency.
\end{itemize}

Deep learning pipelines are memory-bound to a relatively small input size, meaning that it is unfeasible to use the full DM-time transform of the observation as input. Instead, it is necessary to use search windows of the DM-time space of a constant size. Obtaining DM-time windows is a trivial task provided that the (candidate) width, DM, and time are already known. However, an optimal slicing strategy must be devised in the case of a direct search as no prior information on the properties of the signal are known. This strategy must ensure, among other things, that the DM-time plane is sliced into windows of suitable size and overlap to ensure that no candidates are overlooked.

\begin{figure}
	\includegraphics[width=\columnwidth]{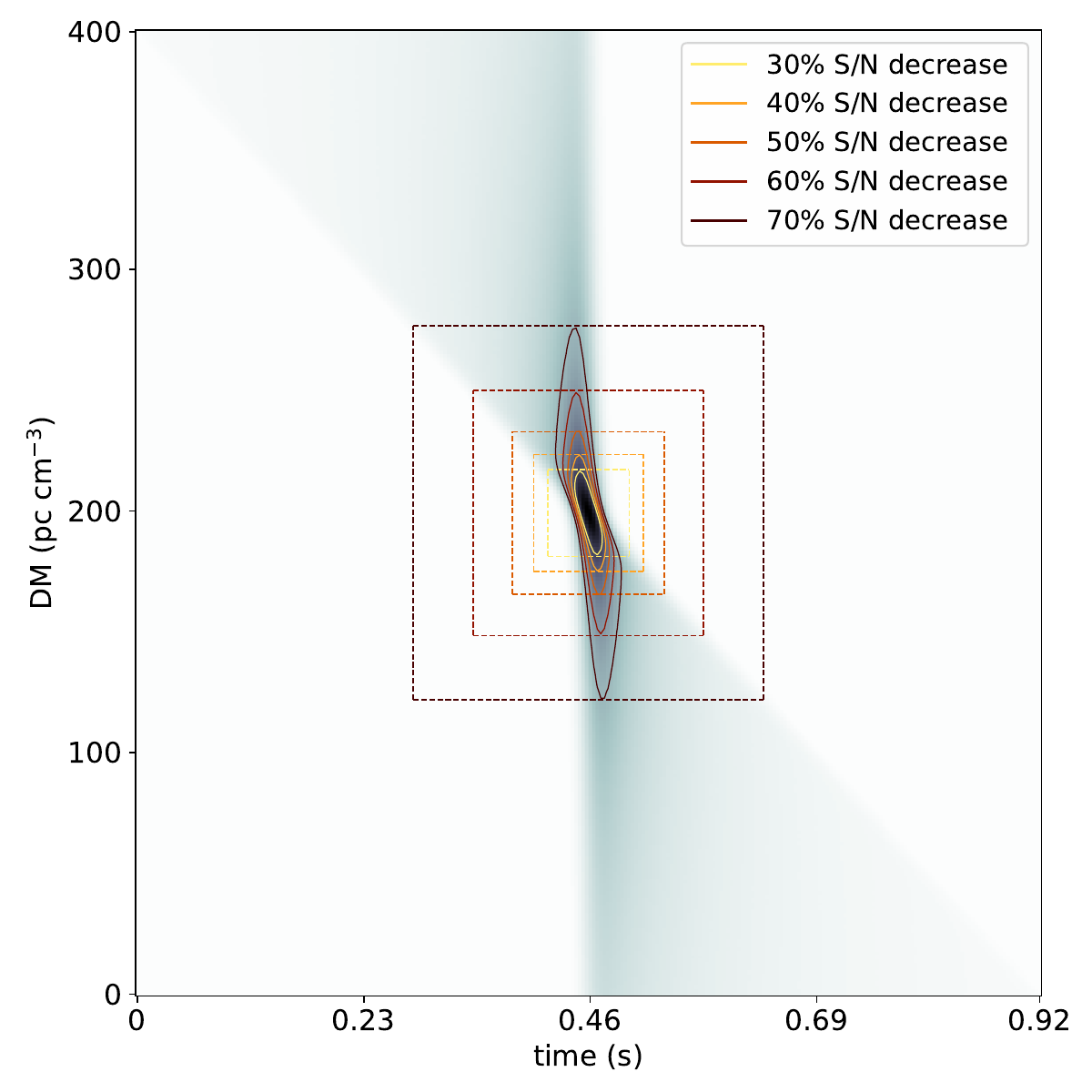}
    \caption{DM-time transform of a simulated burst and contour lines at different S/N drop points. The distance between the furthest point and the central DM-time point can be used as a reference to indicate the minimum amount of signal that needs to be present in the image to be detected by our pipeline, as shown by the dashed line boxes.}
    \label{fig:overlaps}
\end{figure}

To account for the varying aspect ratios and S/N extents that the signal may exhibit, we define the most optimal time ($\Delta t_{\mathrm{range}}$) and DM ($\Delta \mathrm{DM_{range}}$) ranges to slice as
\begin{equation}
    \Delta t_{\mathrm{range}} = \kappa \times w \,,
    \label{eq:optimal_time}
\end{equation}
where $w$ is the width of the burst in seconds and $\kappa$ is a constant related to the amount the S/N has decreased compared to the S/N at the optimal DM (see below); and
\begin{equation}
    \Delta \mathrm{DM_{range}} = \frac{\Delta t_{\mathrm{range}}}{\mathcal{D} \times (\nu_{\mathrm{bottom}}^{-2} - \nu_{\mathrm{top}}^{-2})} \,,
    \label{eq:optimal_DM}
\end{equation}
where $\nu_{\mathrm{bottom}}$ and $\nu_{\mathrm{top}}$ are the radio frequencies at the bottom and top of the band.

The constant $\kappa$ can be related to a S/N decrease due to dedispersing the data at a wrong DM value. \citet{2003ApJ...596.1142C} found a relation between the S/N decrease as a function of DM, pulse width, bandwidth, and observing frequency. The fractional peak S/N decrease for a DM error $\delta$DM is defined as 
\begin{equation} \label{eq:snr_drop_cordes}
    \frac{\textrm{S/N}(\delta \textrm{DM})}{\textrm{S/N}} = \frac{\sqrt{\pi}}{2} \: \zeta^{-1} \, \textrm{erf} \: \zeta \,,
\end{equation}
where $\textrm{erf}$ is the error function and
\begin{equation}
    \label{eq:zeta}
    \zeta = 6.91 \times 10^{-3} \, \delta \textrm{DM} \: \frac{\Delta \nu_{\textrm{MHz}}}{W_{\textrm{ms}} \nu^{3}_{\textrm{GHz}}} \,,
\end{equation}
where $\Delta \nu_{\textrm{MHz}}$ is the bandwidth in MHz, $W_{\textrm{ms}}$ is the burst width in ms and $\nu_{\textrm{GHz}}$ is the observing frequency in GHz. Note that this relation is only valid in the small bandwidth approximation, where $\Delta \nu \ll \nu$. Modern telescopes incorporate larger bandwidths that break this limit. To obtain the S/N decrease at a general configuration, equation \ref{eq:snr_drop_cordes} can be integrated from the bottom to the top frequencies in the band
\begin{equation} \label{eq:cordes_integral}
    \frac{\textrm{S/N}(\delta \textrm{DM})}{\textrm{S/N}} = \int_{\nu_{\textrm{bottom}}}^{\nu_{\textrm{top}}} \frac{\sqrt{\pi}}{2} \: \zeta^{-1} \, \textrm{erf}  \: \zeta \: \textrm{d}\nu_{\textrm{GHz}} \,.
\end{equation}
The constant $\kappa$ from equation \ref{eq:optimal_time} can be equated to the $\delta \textrm{DM}$ from equation \ref{eq:cordes_integral} after substituting in equation \ref{eq:zeta}, since the other variables in $\zeta$ are fixed. Using equations \ref{eq:optimal_time} and \ref{eq:optimal_DM} it is now possible to determine a suitable DM and time range for any given burst width and signal-to-noise degradation level, ensuring uniform aspect ratio and size across all signals, irrespective of their intrinsic features. Since the width of the signal is unknown, multiple window sizes are required to cover a wide range of possible widths in the search. Our method adopts a strategy akin to the traditional matched filter search, employing trial widths that are powers of two. This choice optimises computational efficiency whilst optimising sensitivity to various burst widths.

The next step is to choose a slicing strategy. The challenge in optimally scanning the DM-time space is to find an overlap that would allow bursts at the edges of the window to be detected while generating the fewest windows. If the centre of the bow tie was located at the last time sample of the window and there was no overlap with the next window, our pipeline would struggle to correctly identify the burst since only half of the information would be available. The most optimal overlap is then defined by the minimum amount of signal that needs to be present in the window for the pipeline to detect it. We used the S/N degradation as a function of DM to control the amount of signal present in the search window. By choosing a S/N decrease, contour lines at the given S/N decrease can be used to find the corresponding region around the bow tie centre. A larger allowed S/N decrease probes further into the bow tie structure, increasing the area around the bow tie that will be present. The distance between the centre of the signal to the furthest point in the contour line can be used as an overlap value. Figure \ref{fig:overlaps} shows the area around the centre of the bow tie at different fractional S/N decrease points. At the most optimal DM and time ranges, the signal morphology in the DM-time plane becomes standarised, such that all bursts occupy similarly sized regions, permitting the S/N decrease to dictate the overlap size. In section \ref{section:Results}, we explore the effects of using different overlaps on the performance of the model.

\subsection{Dedispersion strategy}

\begin{figure}
	\includegraphics[width=\columnwidth]{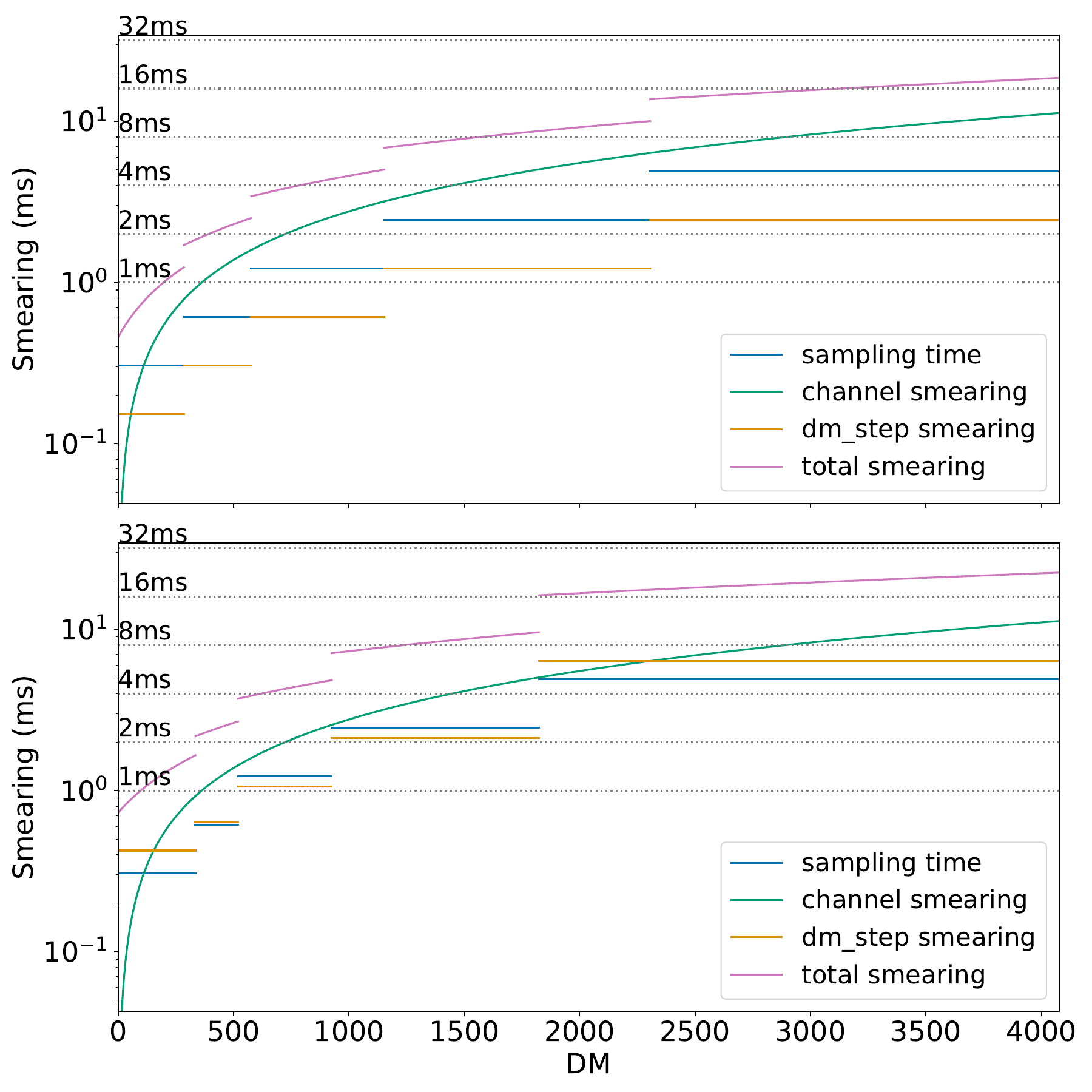}
    \caption{Smearing time as a function of DM for a dedispersion scheme obtained from our pipeline (top) and the DDplan script from PRESTO (bottom). The green, orange, and blue lines show the intrachannel, DM step, and sampling time smearing respectively. The solid purple line indicates the sum of all the smearing effects. The dotted line shows the width values used for the search.}
    \label{fig:ddplans}
\end{figure}

\begin{table}
	\centering
	\caption{Table summarises the dedispersion schedule for an L-band observation with the MeerKAT telescope. The scheme was obtained using our pipeline.}
	\label{tab:our_ddplan}
	\begin{tabular}{lccc} 
		\hline
		Low DM & High DM & DM step & Number of DM trials\\
		\hline
		0.0 & 284.5 & 0.072 & 3946\\
		284.5 & 576.3 & 0.144 & 2023\\
		576.3 & 1152.5 & 0.288 & 1998\\
        1152.5 & 2305.1 & 0.577 & 1998\\
        2305.1 & 4077.1 & 1.154 & 1536\\
		\hline
	\end{tabular}
\end{table}

\begin{table}
	\centering
	\caption{Table summarises the dedispersion schedule for an L-band observation with the MeerKAT telescope. The scheme was obtained using the DDplan from PRESTO.}
	\label{tab:PRESTO_ddplan}
	\begin{tabular}{lccc} 
		\hline
		Low DM & High DM & DM step & Number of DM trials\\
		\hline
		0.0 & 334.4 & 0.20 & 1672\\
		334.4 & 518.9 & 0.30 & 615\\
		518.9 & 925.4 & 0.50 & 813\\
        925.4 & 1822.4 & 1.00 & 897\\
        1822.4 & 4000.4 & 3.00 & 726\\
		\hline
	\end{tabular}
\end{table}

An efficient dedispersion scheme and an optimal choice of DM step are crucial components of a single pulse search, given that incoherent dedispersion of the filterbank data is the most computationally expensive step in the pipeline. A coarse DM step reduces the time and complexity of the dedispersion stage but at the cost of decreased sensitivity, potentially missing some sources. Conversely, a fine DM step increases redundancy in the search, unnecessarily extending the compute time.

In an incoherent dedispersion scheme, there are three time-smearing components to consider. The first is sampling time smearing, which arises from the discrete time resolution at which the data are acquired. The second is intra-channel smearing, an effect that incoherent dedispersion cannot correct for, caused by the discretisation of the filterbank data into independent frequency channels. The intrinsic delay between the top and bottom frequencies of each frequency channel is equal to 
\begin{equation}
\Delta t = 8.3 \times 10^{6} \times \textrm{DM} \times \Delta \nu_{\textrm{channel}} \times \nu^{-3} \textrm{ms} \,,
\label{eq:intrachannel_smearing}
\end{equation}
where both the central frequency $\nu$ and channel bandwidth $\Delta \nu$ are in MHz. The third component is DM step smearing, which relates to the extra smearing arising from dedispersing the signal at a wrong DM due to the finite DM step. It is defined as the total time delay across the bandwidth when the true DM is at half the DM step.

Some dedispersion strategies, e.g. {\sc presto}'s DDplan, define the first DM step as the dispersion measure value that induces a delay across the bandwidth equal to the sampling time smearing. From equation \ref{eq:intrachannel_smearing}, it can be seen that larger DM values lead to increased intra-channel smearing. The so-called diagonal DM is reached when the smearing across a single frequency channel equals the sampling time, indicating the DM values at which intra-channel smearing starts to dominate over other smearing factors. Beyond these values, searching for fine DM steps becomes redundant. Often, at DM values double or triple the diagonal DM, adjacent time samples are added together, increasing the DM step and reducing the compute time of the dedispersion.

Even though the strategy described above is efficient, it cannot be implemented in our pipeline. Dedispersing at the most optimal DM steps within the specified DM ranges results in a DM-time transform with fewer DM trials than the optimal pixel size of the input image for a given $\Delta \mathrm{DM_{range}}$ (see \S\ref{subsec:time_benchmarking}). This would result in DM-time images not being sliced at the most optimal values. We propose a different dedispersion scheme using finer DM steps. The first DM step is chosen the same way as explained above, by finding the DM that would create a smearing across the band equal to the sampling time smearing. Similar to the traditional dedispersion scheme, adjacent time bins are summed at specific DM values, doubling the DM step size. The limiting DMs are defined as the DM values inducing a total smearing that results in the native burst width doubling its original value. This criterion is chosen due to the use of width searches that are powers of two. After the first width is searched, the current DM-time array is downsampled by a factor of two in both DM and time axes, and appended to the next DM-time array in the search. The shape of both DM-time arrays would match since at the limiting DMs the sampling time and DM step are doubled. The next width is then searched, and the process is repeated. This ensures that the full DM range is searched for all widths while the search windows are generated at the optimal DM and time range. The number of DMs trials is modified to accommodate a finite number of images to be sliced at each searching width.

A Python implementation of the dedispersion plan described above has been made publicly available\footnote{\url{https://github.com/sbelmontediaz/Hermes}}. A dedispersion scheme can be determined for any observing configuration. Compared to the traditional method, our strategy uses more DM trials by using finer DM steps. This was done to ensure the DM-time images are extracted at the correct $\Delta \mathrm{DM_{range}}$. Tables \ref{tab:our_ddplan} and \ref{tab:PRESTO_ddplan} show two different dedispersion schemes planned using our approach for MeerKAT L-band observations and the DDplan script from PRESTO. Figure \ref{fig:ddplans} shows the different smearing effects of both dedispersion strategies. Although our proposed dedispersion strategy implies an increase in the number of DM trials to compute, modern GPU-accelerated dedispersion pipelines allow for dedispersion faster than real-time (see a further discussion in \S\ref{subsec:time_benchmarking}).


\section{Deep learning model}

\label{sec:DeepLearning}

\begin{figure*}
	\includegraphics[width=\textwidth]{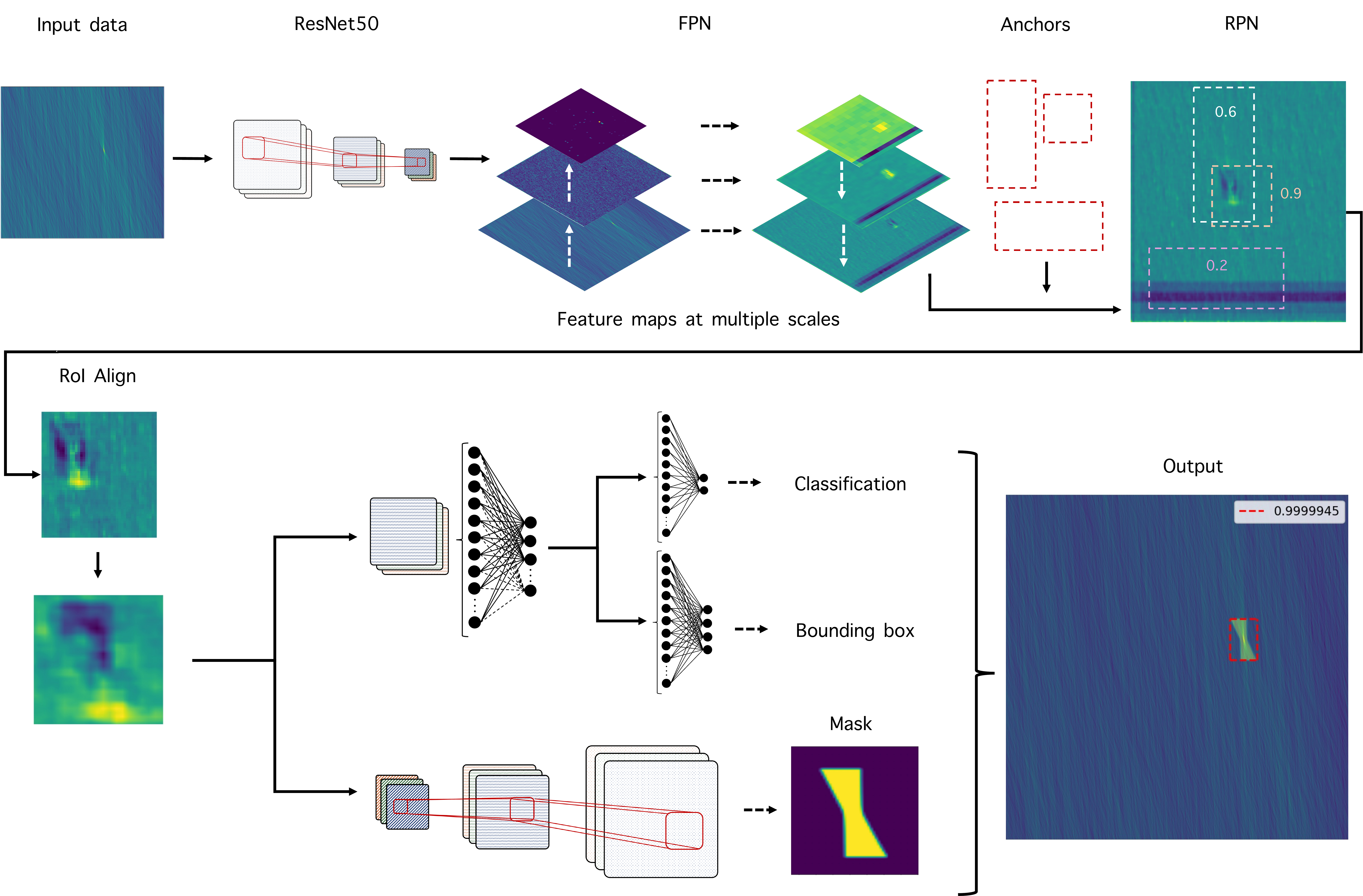}
    \caption{A workflow diagram of our pipeline, showing the classification, localisation and segmentation mask prediction steps. The DM-time transform acts as an input to the algorithm. Feature maps are obtained using a backbone CNN, which are up-sampled by the FPN to obtain high semantics across different scales. The RPN proposes regions of interest across the image where relevant objects may be present, performing a light-weight classification. Regions with the highest score are selected and aligned with the original feature map by the RoI Align network. From these RoI, a segmentation mask is obtained through a CNN with an up-sampling layer at the end, and a classification score and bounding box coordinates are obtained through a series of fully connected layers, producing the output of the algorithm.}
    \label{fig:mask_rcnn_flowchart}
\end{figure*}

Searching for transients in the DM-time transform can be described as a classification task, where real bursts need to be distinguished from RFI and noise. Such a pipeline can benefit from localising the candidate in DM and time to allow further inspection using other diagnostic plots. Therefore, the search is more accurately described as an image segmentation task, in which individual DM and time bins are assessed to identify which belong to a real source and which belong to noise.

Convolutional Neural Networks (CNNs) have been the state-of-the-art in computer vision for the last decade (e.g. \citet{sarraf2021comprehensive}). They are a class of artificial neural networks that are very efficient at extracting local hierarchical features from 2D data. CNNs can be described through three main operations. The input data is first convolved with a set of learnable kernels or filters, in which the kernel is slid across the input data to obtain the dot product between the filter and the overlapping region. At each position, the dot product represents the response of the filter to certain features present in the data, capturing the most relevant information. Therefore, the convolution operation produces a feature map containing the main patterns of interest present in the data. Since the same kernel is applied across different positions, the number of parameters to learn is reduced, improving the efficiency of the network. Moreover, parameter sharing and sparse interactions provide CNNs with the property of translation invariance, since the kernels should be able to capture patterns and features regardless of their spatial location.  After the convolution operation, the feature map is passed through a non-linear activation function, which helps detect non-linear features. Finally, a pooling operation is applied to the non-linear feature map. Pooling returns the summary statistics of a neighbouring region using sliding kernels. The maximum pooling operation is usually chosen, in which the kernel returns the value of the maximum pixel. Pooling reduces the dimensions of the feature map while capturing the more relevant features present in the data, hence increasing the efficiency of the network. These three operations can be applied multiple times. In each convolution step, kernels are weighted to be more sensitive to certain patterns. This allows CNNs to capture local features hierarchically, where the first kernels are more sensitive to general and broad features, such as edges and colours, and the last layers learn to capture smaller and more abstract features.

There is a plethora of image segmentation algorithms based on CNNs to choose from. We considered that a Mask-RCNN model \citep{He_2017_ICCV} was the most suitable option to start with \citep{2021PhDTZafiirah}. The model is based on the Faster-RCNN algorithm, an object detection pipeline that returns the class-predicted score and a bounding box surrounding the object \citep{Girshick_2015_ICCV}. This family of algorithms belongs to the two-stage object detection models, where region proposals are initially generated before subsequent refinement and object classification. This is contrary to the one-stage models, which leverage processing time by directly performing object detection on the data. Faster-RCNN starts by extracting a feature map from the input data using a pre-trained backbone network. The region proposal network (RPN) uses the feature map as an input and outputs a set of rectangular bounding boxes and their corresponding objectness score, which relates to the probability that the region contains an object of interest. This is done by sliding a small network across the feature map. At each location, a set of predefined anchor boxes is generated at different scales and aspect ratios. These act as reference points for the small network. Each region of the feature map covered by the anchor box is mapped to a lower-dimensional representation, reducing memory requirements and computer complexity. The low-dimensional feature vector is passed into two fully connected layers, which perform bounding box regression and binary classification, generating the object proposals and the objectness score. At this stage, multiple bounding boxes can be generated for the same object. Non-maximum suppression (NMS) is used to reduce the overlapping regions and refine the bounding boxes. NMS uses the objectness score to select the region with the highest probability of containing an object. The intersection over union between the bounding boxes is calculated with respect to the reference region. Those with a value larger than a given threshold are considered to be overlapping, for which only the region with the highest score is retained. The second stage of the Faster-RCNN model starts by doing a region of interest (RoI) pooling. Since the proposed regions of interest have different scales and aspect ratios, they need to be normalised to a fixed size and shape to be fed to the next layers. RoI pooling works by dividing the region of interest into a grid of sub-windows. The specific distribution and size of the grid and sub-windows depend on the desired output shape of the RoIs. On each sub-window, a maximum pooling operation is applied to reduce the dimensionality of the RoI to the desired fixed shape while preserving spatial information. The output RoI is passed through a fully connected layer returning a RoI feature vector, which is fed into a set of two fully connected layers that perform bounding box regression and multi-class classification.

Mask-RCNN follows a similar architecture to Faster-RCNN, but it adds two extra networks: a Feature Pyramid Network (FPN) and a segmentation mask regression network. FPNs were designed to address the challenge of detecting multi-scale objects. They use feature maps from a backbone CNN to build a feature pyramid. The original feature maps have different spatial resolutions as the convolution layers downsample the data. The last output of the backbone CNN is the feature map with the lowest spatial resolution but the highest semantic content. To obtain high-level semantic features at various scales, FPNs introduce a top-down pathway by repeatedly upsampling the lowest-resolution feature map. At each resolution level of the top-down pathway, lateral connections are used to merge the feature maps from the original backbone CNN with the upsampled ones, creating a fusion between the high-level semantics features from the top-down pathway and the fine-grained details of the backbone network. The RPN is then independently applied at each different spatial resolution level to generate multi-scale region proposals. The segmentation mask branch is added in parallel to the two fully connected layers performing bounding box regression and object classification after the RoI pooling, and it consists of a series of convolution layers with a deconvolution layer that upsamples the RoI until the segmentation mask is obtained. In order to obtain the segmentation mask, a pixel-by-pixel classification is performed. Hence, it is important to reduce any possible misalignments between the feature map and the RoI. RoI pooling quantises a floating-point RoI and performs a maximum pooling operation to obtain the fixed shape RoI. This introduces misalignments between the RoI and the feature map grid. Mask R-CNN solves this issue by replacing the RoI pooling operation with a RoI align step. Rather than quantising the floating-point RoI, RoI align uses bilinear interpolation to compute the exact values of the feature map at the fractional coordinates of the RoI bins.

The classification, boundary box, and mask branches are computed in parallel, which decouples the classification and the mask prediction. This results in a mask prediction that is independent of the classification, and masks from different classes do not compete. Figure \ref{fig:mask_rcnn_flowchart} shows a flowchart of the Mask R-CNN model. For each computed RoI, the loss function is defined as a multi-task loss by adding the individual losses of each branch. The classification branch uses a categorical cross-entropy loss \citep{Michelucci2019}, whereas the smooth L1 loss \citep{Girshick_2015_ICCV} and the binary cross-entropy loss are used by the bounding box and mask prediction branches. The latter is applied pixel-by-pixel across the segmentation mask.

\begin{figure}
	\includegraphics[width=\columnwidth]{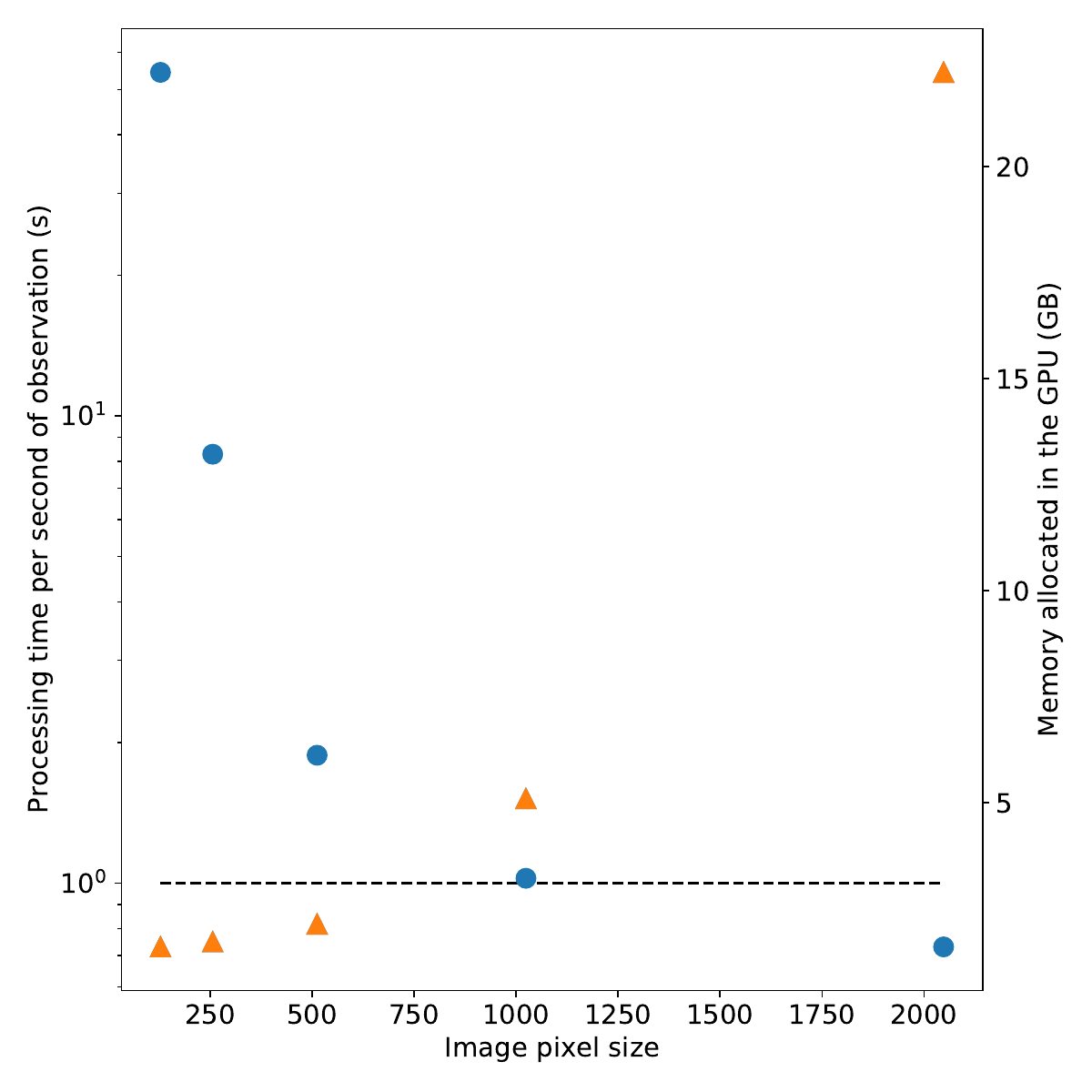}
    \caption{Inference time of the Mask R-CNN model for one second of filterbank data across all DM trials (dots) and memory allocated in the GPU (triangles) shown for different input image sizes. The dashed line shows the boundary for real-time processing.}
    \label{fig:processing_time}
\end{figure}

\subsection{Dataset generation}

\begin{table}
	\centering
	\caption{Table showing the distributions and ranges used to simulate the bursts.}
	\label{tab:properties_bursts}
	\begin{tabular}{lcc} 
		\hline
		Burst property & Distribution & Range\\
		\hline
		DM & uniform & 50 - 4000 pc cm$^{-3}$\\
		Width & uniform & 0.5 - 32 ms\\
		Amplitude & $\chi^{2}(N_{\textrm{d.o.f}}=2)$ &$\mu = 1.4, \sigma = 0.4$\\
        S/N & $\chi^{2}(N_{\textrm{d.o.f}}=2)$ & > 8\\
		\hline
	\end{tabular}
\end{table}

The Mask-RCNN model includes more than 40 million trainable parameters for a ResNet50 backbone \citep{He_2016_CVPR}. In order to facilitate the robust training of the network, it is necessary to provide a substantially large dataset. Therefore, utilising real observations to build a suitable dataset poses a considerable challenge. Although more than 3,500 pulsars have been detected (from the ATNF pulsar catalogue\footnote{\url{https://www.atnf.csiro.au/people/pulsar/psrcat/}} \citet{2005AJ....129.1993M}), only a small fraction of them emit with sufficient intensity to discern individual bursts. Single bursts from nearby pulsars are often detected, offering potential for augmenting the dataset. However, a suitable dataset should include examples that adequately sample the entirety of the parameter space within the underlying distribution. In our case, the dataset should encompass a wide range of DMs and widths. Therefore, incorporating self-similar single bursts from a select few sources would bias the dataset towards the properties observed in nearby pulsars, which does not encompass the whole population of radio transients (for example, FRBs are detected at larger DM values due to their extragalactic origin). 

\begin{figure*}
	\includegraphics[width=\textwidth]{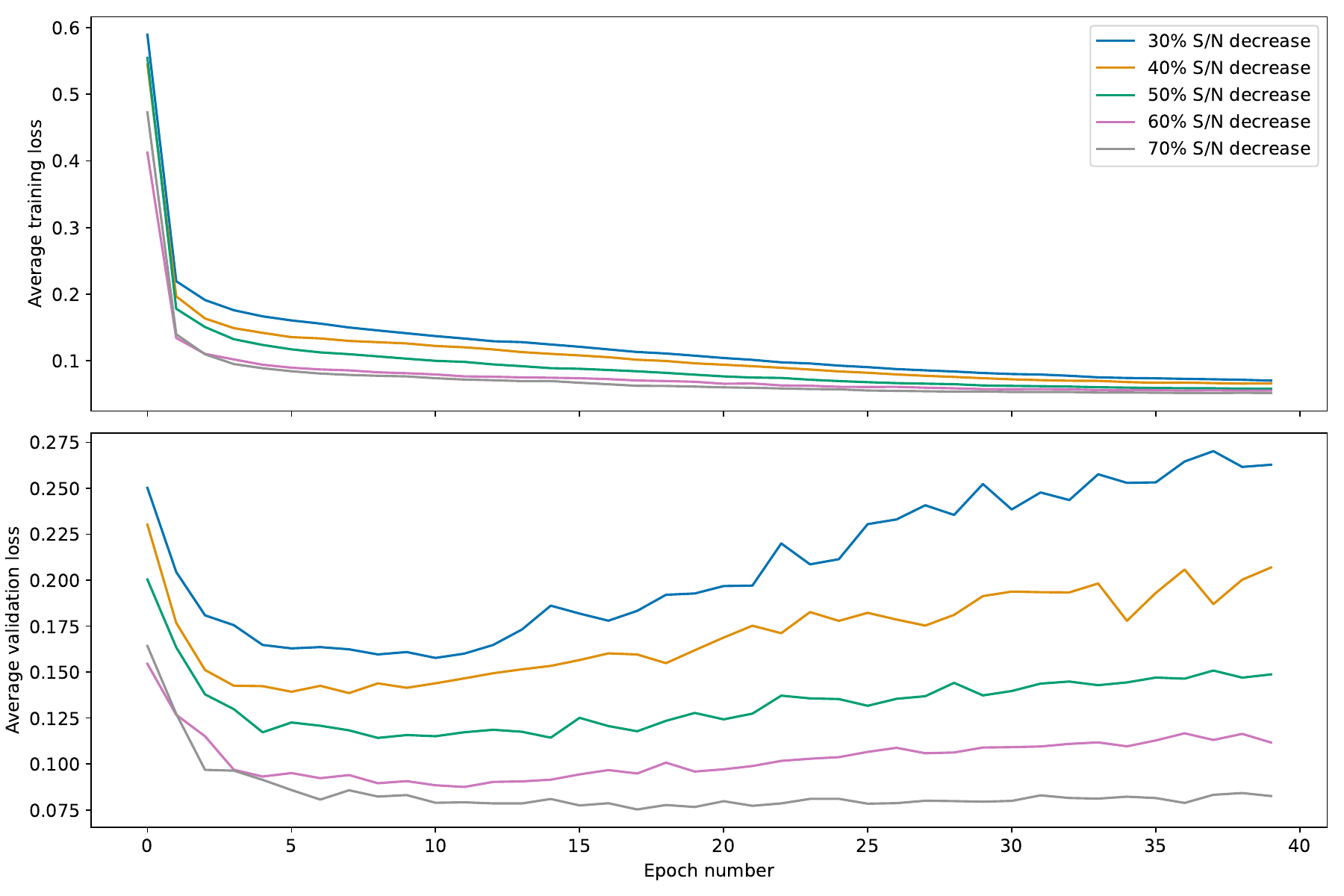}
    \caption{Training (left) and validation (right) loss curves for a training loop of 40 epochs using different mask sizes, where each corresponds to the contour size at a specific S/N drop threshold.}
    \label{fig:train_val_loss}
\end{figure*}

For all the above reasons, we decided to build our dataset through simulations. This approach afforded us complete control over the sampled parameters and the quantity of examples. The bursts were simulated assuming a simple single Gaussian profile that was dispersed to follow a frequency-squared time delay provided a DM value. The bursts were broadband, occupying the full band, and they had a flat power spectrum. No extra effects such as scattering or scintillation were included in the simulations. This simplistic approach enables us to test the basic idea of our model, with plans to add more complexity in future work. The bursts were injected on top of MeerKAT L-band ($\nu_{\rm{centre}} =$ 1284 MHz and $\Delta \nu =$ 856 MHz) observations with a time resolution of 306.24 $\mu$s and 1024 frequency channels, that only contained noise and RFI. These were used to include real noise statistics and to provide examples of real RFI to our dataset. Five files containing ten minutes of radio observations with no detected sources in them were used. Bursts were systematically injected into the noise observations at random time samples one at a time. The DM-time window containing the burst would be saved, and a new burst would be injected into the original noise data, ensuring the absence of concurrent bursts within the dataset. Dedispersion was applied using the corresponding DM step size prescribed by the dedispersion strategy. Subsequently, the DM-time plane was sliced at the most optimal values given the burst width. The search window was downsampled to the desired input pixel size of the Mask-RCNN. The location of the bursts in the image was randomised to ensure the algorithm learns to find targets across the image. The properties of the bursts were sampled following the parameter space of observed transient phenomena and can be seen in Table \ref{tab:properties_bursts}. These values are based on the MeerKAT L-band filterbank\footnote{Filterbank files are in sigproc \citep{sigproc_report} format containing frequency, time, and amplitude.} observing configuration, which limits the minimum width of the bursts.

In total, the dataset contains 23,000 examples. The training, validation, and test split are 10,000, 2,000, and 11,000 respectively. One thousand of the test examples are images with no sources injected in them, which enables benchmarking on the noise classification. The labels are JSON files containing the coordinates of the segmentation mask and the classification label. The normalisation process entailed calculating the sigma-clipped mean and standard deviation. Pixel values were scaled to ensure that the lowest pixel value represented the mean minus four times the standard deviation, and the highest pixel value represented the mean plus eight times the standard deviation, effectively distributing the data within a dynamic range from 0 to 255. The number of image channels was increased to three by duplicating the data from the original channel, where in this context, `channels' refers to the Red, Green, and Blue (RGB) color channels of an image. These steps were necessary due to the use of pre-trained weights from the COCO dataset \citep{COCO}, which contains three-channel RGB JPEG images.

\subsection{Training}

The Mask-RCNN module from the torchvision library from pytorch was used\footnote{\url{https://pytorch.org/vision/main/models/mask_rcnn.html}}. \textsc{pytorch} provides full pre-trained COCO weights for a Mask-RCNN network using a ResNet50 backbone.
The model's hyperparameters were kept at the default values provided by the torchvision model, with the exception of the minimum and maximum image sizes. These values were set to match the pixel dimensions of the dataset (see \S\ref{subsec:time_benchmarking}). Fine tuning of the model's hyperparameters could yield better results. However, the starting setup gives a good performance (see \S\ref{subsec:mask_size_effect}). The training was started from the pre-trained COCO weights, following a transfer learning strategy. The optimiser used was Adam. A constant learning rate of $10^{-5}$ and a batch size of 2 was used. The network was trained for 40 epochs. This was a large enough number to see a constant increase in validation loss, marking the point at which the model stopped learning general features from the data. An NVIDIA RTX-3090 GPU with 24GB of memory was used to train the model. 

\section{Results and Analysis}
\label{section:Results}
\subsection{Time benchmarking}
\label{subsec:time_benchmarking}

Radio transient searching pipelines need to run in real time. The vast amount of data recorded by modern telescopes makes off-line searches impractical. Moreover, real time detection of transient events enables multi-wavelength follow-up observations, providing valuable insights into the nature of these events. Our proposed pipeline can be divided into four processing blocks that can run in parallel: dedispersion, DM-time slicing and normalisation, Mask-RCNN inference, and results collection. Dedispersion is often the bottleneck of searching pipelines. However, GPU-accelerated algorithms achieve faster than real-time dedispersion (e.g. \citet{2020ApJS..247...56A}). The pre-processing and post-processing of the data (slicing and result collection) were not fully optimised for real-time performance. The focus was put on optimising the inference time of the neural network since it can be considered the main step of the pipeline. It is bound by the model size, hardware and input image size, which affects the total number of images to process. A Mask-RCNN model with a ResNet50 as a backbone network has more than 40 million parameters. Fine tuning of the image size plays a key role to achieve real time performance. On one hand, inference time of smaller images is reduced compared to larger sizes, suggesting that reducing the input size would reduce the overall processing time. On the other hand, due to the overlap applied in our slicing, using a smaller input size increases the total number of images to process. From Figure \ref{fig:processing_time}, it is clear that the reduction in image size has a larger effect than the increase in inference time per image, and that real time performance can be reached by increasing the input size. 

The downside of working with larger images is the increase in memory allocated in the GPU. Increasing the image size to 2048x2048 pixels would allow faster than real time processing, but the memory allocated in the GPU would be larger than 90 $\%$ of the hardware used. The values presented in Figure \ref{fig:processing_time} were obtained during inference, where only the forward pass is executed, and gradient information is not retained, as no backpropagation is required. In contrast, training the model would demand additional memory to store gradients and other intermediate values. Thus, given our hardware, we decided to work with images of size 1024x1024 pixels. This choice ensures near real time inference performance without substantially increasing the GPU's memory load.

It is also important to note that the number of images to process is also controlled by the configuration of the observation. The sampling time, bandwidth, and number of frequency channels will produce different dedispersion plans with more or less DM samples. Other user-defined parameters also affect the number of images to process, such as the DM range to search for or the number and sizes of burst widths. The numbers provided in Figure \ref{fig:processing_time} were obtained from processing a $\sim$55s long MeerKAT observation in L-band ($\nu_{\rm{centre}} =$ 1284 MHz and $\Delta \nu =$ 856 MHz), 1024 frequency channels and a sampling time of $\sim$ 306 $\mu$s, where the DM range searched was 0 pc cm$^{-3}$ -- 4,000 pc cm$^{-3}$ and the range of burst width searched was 1 ms -- 32 ms, generating a total of 1,438 images which were processed in $\sim$56s.

\begin{figure}
	\includegraphics[width=\columnwidth]{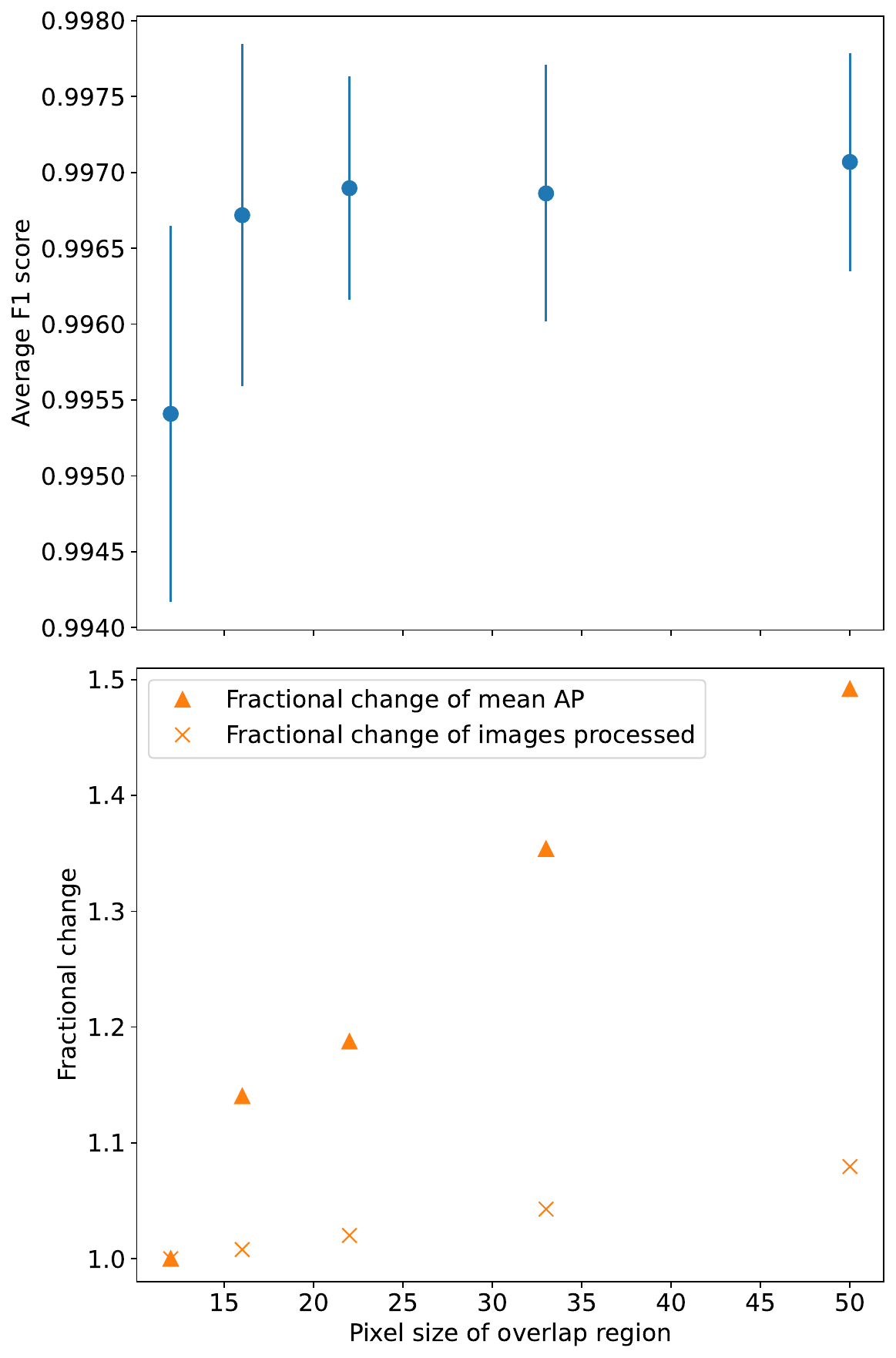}
    \caption{Performance and efficiency metrics of the model for different overlap sizes. The overlap region is directly related to the mask size, where the smallest overlap region corresponds to the highest S/N drop point. On the top is the average F1 score. The test set was divided into 10 splits, and the F1 score was calculated for the different models over the last 35 epochs of training. The sigma-clipped mean and standard deviation were computed for each split, with the plotted values corresponding to the mean of these statistics. On the bottom, the triangles indicate the fractional change in average AP across different IoUs and crosses represent the fractional change in the number of images to be processed. The values for the smallest mask size are used as the reference.}
    \label{fig:clipped_f1}
\end{figure}

\begin{figure*}
	\includegraphics[width=\textwidth]{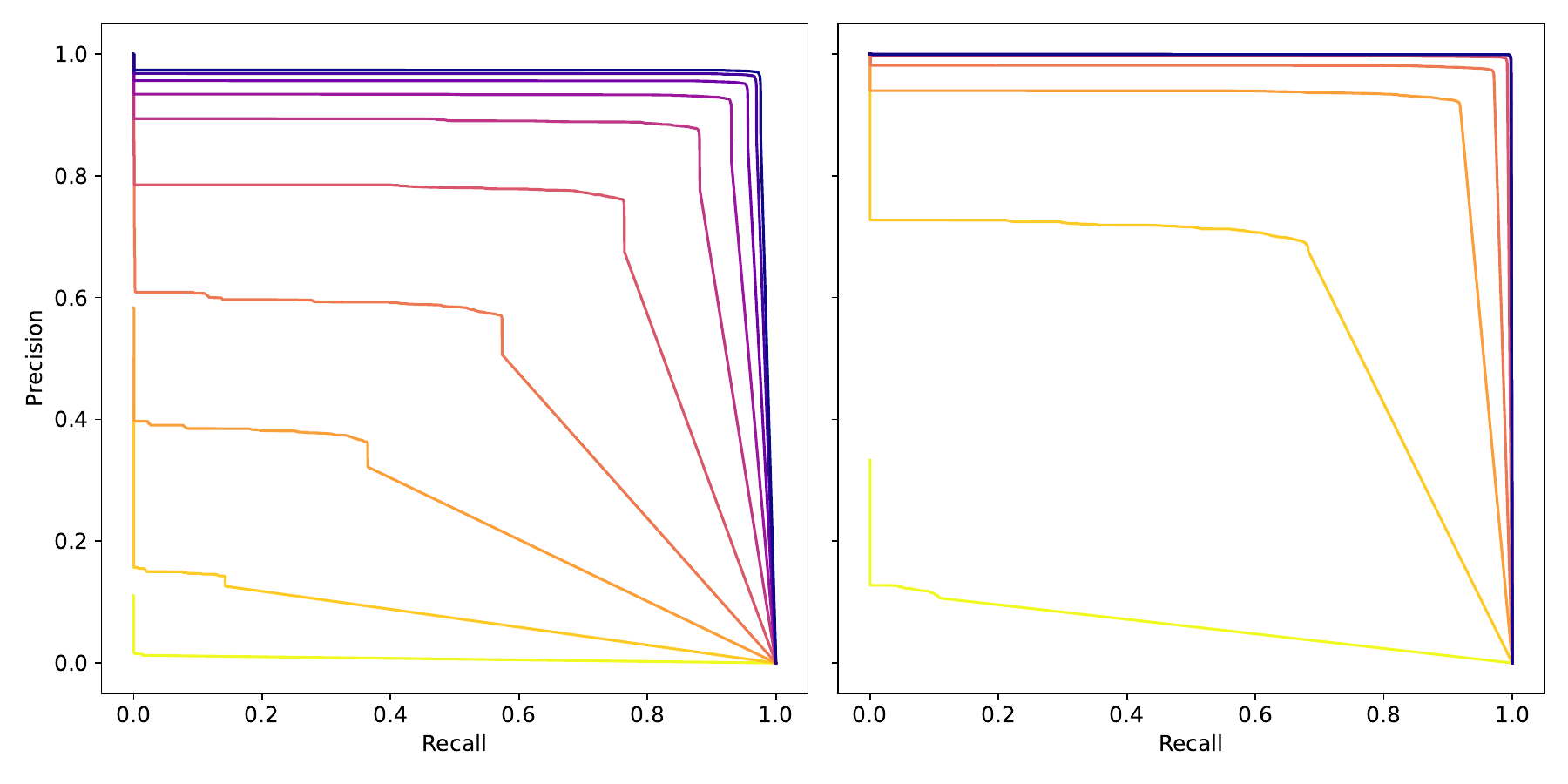}
    \caption{Precision-recall curves at different IoU thresholds for a mask size corresponding to a S/N drop of 70$\%$ (left) and 30$\%$ (right). Curves correspond to an IoU of 0.95 for the yellow, lowering the threshold to the purple colored lines in steps of 0.05, reaching a minimum threshold of 0.5. The APs corresponding to the smaller mask size range from 0.021 to 0.950 from the highest to lowest IoU threshold, and 0.490 to 0.997 for the respective IoU threshold for the largest mask size.}
    \label{fig:APs}
\end{figure*}

\subsection{Impact of mask size on model performance}
\label{subsec:mask_size_effect}
In our proposed pipeline, the size of the mask emerged as a critical factor specific to our task. Typically, mask sizes are left flexible to accommodate varying object sizes and viewing angles. However, in our work, different window sizes are used to keep the signal size constant, meaning that the mask size is a new hyperparameter to consider. To study its effect, the model was trained using different masks and overlap sizes. Since the overlap is defined by the minimum amount of signal required to identify a source in the image, it is closely related to the target segmentation mask. As discussed in section \ref{sec:Slicing}, the S/N decrease, caused by dedispersing the signal at the incorrect DM, was utilised to determine the pixel size of the overlaps (see Figure \ref{fig:overlaps}). Using bigger overlaps requires processing more images, increasing the compute time of the pipeline. However, using a bigger mask may improve the performance of the model since more information on the S/N degradation is used to identify candidates. Figure \ref{fig:train_val_loss} shows the training and validation loss for the different masks and overlaps. The training and validation losses are lower for larger masks (corresponding to a larger S/N decrease), indicating improved model predictions. The validation loss curves show that models trained with smaller masks exhibit overfitting at earlier epochs, whereas the model trained with the largest mask does not demonstrate any overfitting. This suggests that larger masks enhance the model's ability to generalise the information from the training set.

To benchmark the effect of the mask size on unseen data, we first used the F1-score statistic. The F1-score is defined as the harmonic mean of precision and recall, where precision measures the accuracy of the positive predictions and the recall measures the classifier's ability to find all relevant instances. Hence, the F1-score is a measure of the balance between precision and recall in a classifier. The test set was randomly divided into 10 groups, each containing 1,000 examples. For each group, the F1-score was obtained for the model trained in the last 35 epochs. The sigma-clipped mean and standard deviation of the F1-scores were calculated for each group. Sigma-clipped statistics were used because, for certain epochs, performance dropped significantly compared to the overall trend. Figure \ref{fig:clipped_f1} shows the averages of the sigma-clipped means and sigma-clipped standard deviations for each mask size. Performance appears relatively consistent across all mask sizes, with only a slight increase in F1 score observed for larger masks. Although a minor improvement is seen with increasing mask size, the differences fall within a very narrow range, and the standard deviations suggest that the F1 scores are consistent with being constant across different mask sizes.

However, the F1-score benchmarks the classification part of the pipeline, but it does not include any information on object detection. To benchmark an image segmentation algorithm, the metric used is the Average Precision (AP), which corresponds to the area under the precision versus recall curve \citep{AP}. The AP is measured by obtaining the precision and recall at different prediction thresholds. Predictions with a very high confidence score are likely to be true positives, maximising the precision. However, setting a very high threshold could potentially miss some less confident positive predictions, minimising the recall. As the prediction score is decreased, the number of false positives increases while the number of false negatives decreases, reaching the highest recall and lowest precision at the lowest prediction score. A perfect classifier would show a precision-recall curve that would stay at a high precision value for most recalls, only dropping down when approaching a recall value of unity. Precision-recall curves can be obtained at different Intersection over Union (IoU) thresholds. The IoU is the ratio between the intersecting area of the predicted mask and the ground truth mask and the sum of the areas of both masks. An IoU of 1.0 would correspond to a perfect alignment between the ground truth and the prediction, and an IoU of 0.0 corresponds to a complete misalignment. The IoU can be used to set a threshold over which the prediction is considered to be a true positive or a false positive. Obtaining precision-recall curves at different IoU levels allows us to obtain more in-depth information on the object detection performance. Figure \ref{fig:APs} shows precision-recall curves at different IoU thresholds from 0.5 to 0.9 for the smallest and largest mask. Across all IoU thresholds, the largest mask consistently outperforms the smallest, with the difference in performance especially pronounced at higher IoU values. While the curves for the largest mask remain close to the top-right corner -- indicating high precision and recall -- the smallest mask exhibits a notable drop in performance as the IoU threshold increases. This trend is observed across all mask sizes: as the mask size increases, the precision-recall curves shift toward the top-right, reflecting improved detection and segmentation accuracy. At the largest mask size, the model achieves excellent performance for IoU thresholds below 0.85, with precision–recall curves tightly clustered in the top-right region of Figure~\ref{fig:APs}, where the AP exceeds 0.95

The limitation of increasing the mask size lies in the increase in compute time. As discussed previously, the mask size is closely related to the overlap used when slicing the DM-time array. Expanding the mask size results in an increased overlap, which results in needing to process a greater number of windows. The fractional increase $R(\alpha_{2}/\alpha_{1})$ in the number of images at two different overlaps $\alpha_{2}$ and $\alpha_{1}$ can be expressed as 
\begin{equation}
    R(\alpha_{2}/\alpha_{1}) = \frac{(w_{x}-\alpha_{1})(w_{y}-\alpha_{1})}{(w_{x}-\alpha_{2})(w_{y}-\alpha_{2})} \,,
    \label{eq:fractional_image_increase_overlap}
\end{equation}
where $w_{x}$ and $w_{y}$ are the window size in the x and y direction respectively. To compare the fractional change in the models' performance and the fractional increase in images processed at different mask sizes, the average AP across all IoUs was used as a statistic to measure the model's performance. Figure \ref{fig:clipped_f1} shows the fractional change in the average AP and the fractional image increase for different mask sizes. From the smallest to the largest mask size, the increase in the number of images to process is 8$\%$, whereas the increase in the mean AP across different mask sizes is 50$\%$. As the increase in performance is much larger than the increase in processing time, we decided to use the largest mask model.

\subsection{Performance on real data}

\begin{figure}
	\includegraphics[width=\columnwidth]{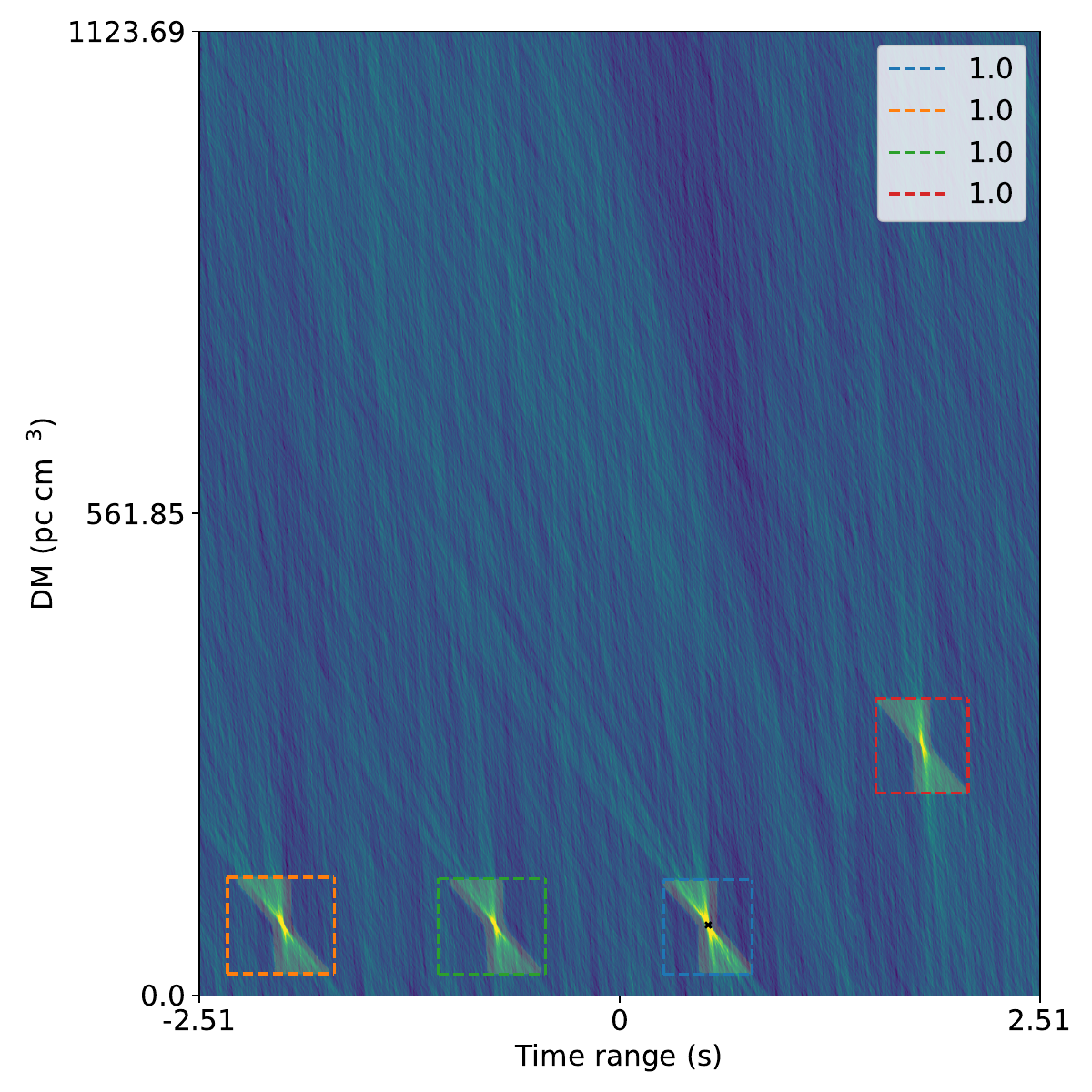}
    \caption{DM - time transform plot showing the output of our pipeline for an image containing four real bursts from nearby, bright pulsars. Three bursts are from the same pulsar, hence located at the same DM. The last burst corresponds to a different pulsar at a different DM. Our pipeline correctly identifies and localises the four bursts, returning a boundary box, segmentation mask, and prediction score for each source. The prediction scores are all unity, corresponding to a high confidence detection.}
    \label{fig:multiple_bursts}
\end{figure}

The model shows good performance on the simulated test set. However, these were relatively simple simulations of single Gaussian component bursts with no exotic features that would distort the signal shape in DM-time space. Before using the model in a real time transient search, it is important to check whether it generalises well to real data that show more complex features. Therefore, a set of MeerKAT filterbank files containing real observations of known bright pulsars and Fast Radio Burst events at S-band, L-band and UHF were used to test our pipeline. The test sample corresponds to bursts of DMs ranging from 3 pc cm$^{-3}$ to 2270 pc cm$^{-3}$ and a diversity of S/Ns and burst widths. All of the four FRBs were successfully detected. Out of the 107 single pulses from a variety of pulsars, 106 were found. This shows that our model is agnostic to observing configuration, since bursts were detected in three different frequency bands. The missed burst was very badly affected by the zero-DM matched filter, which tries to remove RFI by subtracting a zero-dispersion signal template from each frequency channel \citep{2019Men}. 

We note that our pipeline is also able to detect multiple individual bursts in the same window, as shown in Figure \ref{fig:multiple_bursts}, where not only are the three bursts at the same DM location detected, but also at a higher DM of another pulsar \footnote{The filterbank used here was artificially made by appending different files containing different pulsars together.}. This shows the ability of our model to detect bursts anywhere in the image, regardless of their DM and time. The ability to find multiple bursts in the same search window is something previous ML classifiers could not do, since the input image was always centred around the candidate burst, and shows a promising application for being able to detect multiple component bursts. We should stress that this was achieved without training our Mask-RCNN specifically to recognise multiple burst in single images. Only four false positive candidates were identified and by inspection they were all due to residual RFI present in the data.

\begin{figure}
	\includegraphics[width=\columnwidth]{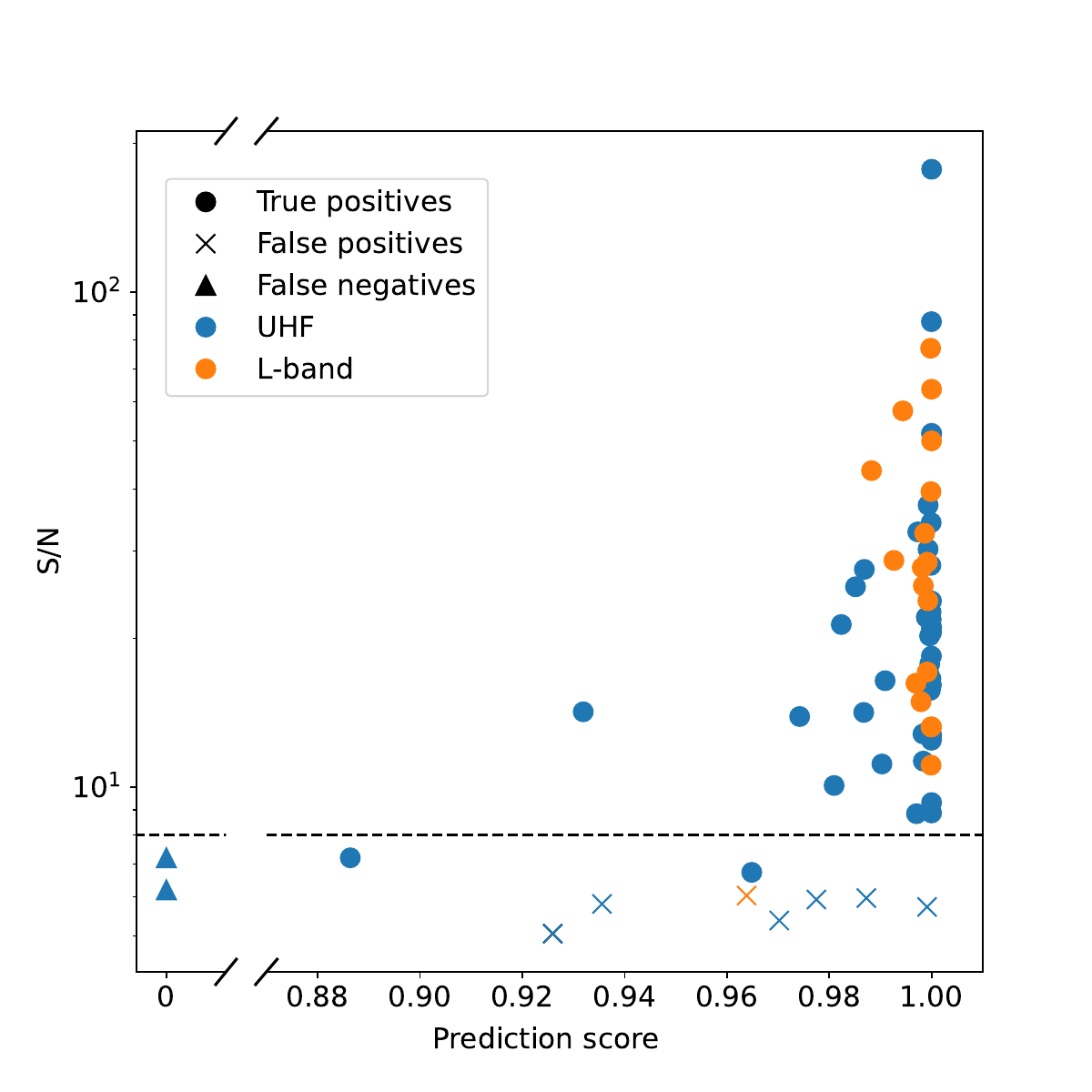}
    \caption{Summary of the matched filtered S/Ns vs prediction score returned by our pipeline for the 64 bursts detected by MeerKAT from the repeater FRB20240114A. The S/N reported are those returned from candreport. The dots show true positives, crosses false positives, and triangles false negatives. The orange colour shows bursts found at L-band and the blue colour represents bursts found at UHF. The corresponding prediction score of the false negatives is zero, note the broken axes to improve the readability of the figure. The horizontal dashed line shows a S/N threshold of eight, which is commonly used in transient searches. Note the log scale in the S/N axis.}
    \label{fig:FRB20240114A_SNR}
\end{figure}

\begin{figure}
	\includegraphics[width=\columnwidth]{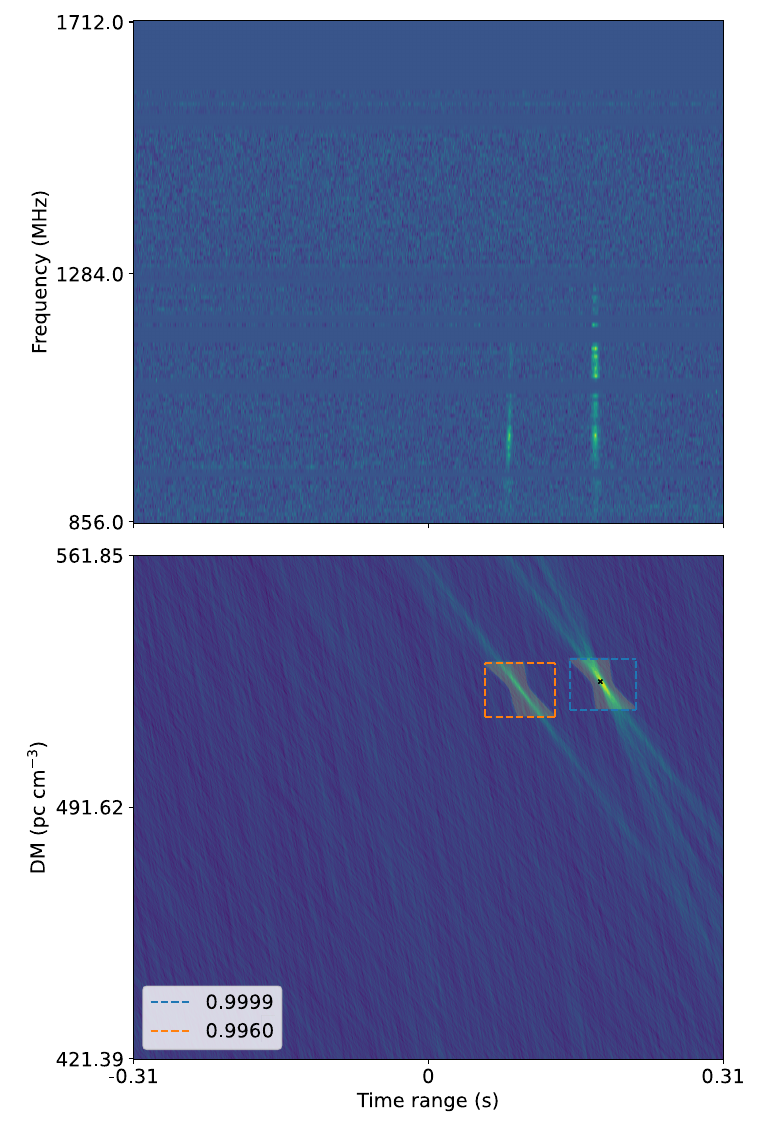}
    \caption{Plots showing two bursts from FRB20240114A from \citet{2024MNRAS.tmp.1973T} at L-band detected by our pipeline. The top plot shows the dedispersed frequency-time data and the bottom shows the DM-time transform with the output from our pipeline. The time axis of both plots is aligned. The frequency-time data was dedispersed at the DM that maximises the S/N of the second burst.}
    \label{fig:multiple_bursts_FRB2024}
\end{figure}

\citet{2024ATel16420....1S} reported a CHIME detection of FRB20240114A, a repeater that was thought to have a high burst rate. The MeerTRAP collaboration (More Transients and Pulsars; \citet{2018MeerTRAP}, \citet{2024MeerTRAP}) conducted follow-up observations with the MeerKAT radio telescope in L-band and UHF ($\nu_{\rm{centre}} =$ 816 MHz and $\Delta \nu =$ 272 MHz).  62 bursts were detected, 18 in L-band and 44 in UHF \citep{2024MNRAS.tmp.1973T}. These observations constitute a very valuable test set since emission from FRB repeaters usually show exotic phenomena such as narrowband bursts, a diversity in their spectra, and "sad-trombone" frequency drifting effects. Figure \ref{fig:FRB20240114A_SNR} shows a summary of the matched filtered S/Ns vs the prediction score returned by our pipeline. Out of the 62 bursts, 60 were correctly identified. Most bursts have an associated high score, 80$\%$ and $94\%$ of the bursts in UHF and L-band respectively were identified with a prediction score > 0.99. This also shows an interesting feature of the model: the prediction score does not depend solely on the S/N of the bursts. Some pulses with a S/N close to the threshold value were identified with a higher score than brighter events. Further manual inspection seems to indicate that bursts with lower prediction scores have shapes in DM-time that deviate further from the ideal shape expected from a broadband emission. The two missed events were burst components too close to the main pulse event for the searching box to fit both, resulting in only the main burst being detected. Specifically, the separation between both events is less than 90ms for both false negatives. Figure \ref{fig:multiple_bursts_FRB2024} demonstrates that when the separation is sufficient, individual components can be detected. This occurs because the non-maximum suppression applied by the Mask-RCNN retains only the prediction with the highest prediction score for overlapping ROIs. Additionally, the two FNs were faint bursts with a S/N below the commonly used threshold of eight, so they were also missed by the thresholding method in the original search and only found later in a visual inspection of the data.

Notably, our model showed that it could potentially detect these bursts if they were isolated, as it detected two other bursts with an S/N below the threshold. These bursts would have been missed in a traditional search pipeline, as they were also found only when manually inspecting the filterbank files to find faint bursts. In fact, one of these bursts was initially missed during manual inspection and was only added to the record after our pipeline identified it. This underscores the advantage of an informed transient search over a brute-force thresholding approach, highlighting the potential of our method to significantly improve radio transient detection. Eight search windows returned false positives. All the FPs corresponded to random noise fluctuations that exhibited what looked like a small dispersion relation, partially mimicking the bow tie shape expected from real astrophysical sources. In total, 7029 images were searched, resulting in an FP rate of 0.1 $\%$. While this result is promising, caution should be exercised in drawing conclusions from this number. The filterbank files used for the search all contain a real source, which could bias the FP rate. A more comprehensive analysis of the FP rate of our pipeline is beyond the scope of this paper.

A final remark is that these bursts exhibited features not present in our test set. As shown in Figure \ref{fig:multiple_bursts_FRB2024}, the bursts are narrowband and have a spectrum that deviates from the flat spectrum used in our simulations. These two effects distort the signal from the ideal bow tie shape in the DM-time space. Nevertheless, the model was resilient enough to successfully identify all of these bursts. The segmentation mask returned represents the shape of the signal had it been broadband. Moreover, both bursts have slightly different S/N maximising DMs, but the model can correctly identify both of them at their corresponding DMs. This demonstrates the significant advantage of using large deep-learning models to address complex tasks.

\section{Conclusions}
In this work, we have presented a novel end-to-end deep learning pipeline for the detection of fast radio transients. Traditional thresholding methods do not fully exploit all available information when selecting candidates. Our method directly searches for transients in the DM–time plane, where astrophysical signals exhibit a signal-to-noise (S/N) degradation characteristic of a bow tie shape.

As the shape and extent of the S/N degradation depend on the number of DM and time bins selected, as well as the intrinsic burst width, we defined optimal DM and time window sizes to preserve the signal morphology. These windows are burst-width dependent, necessitating the use of multiple search widths.

We employed a Mask R-CNN, a deep learning model widely adopted in object detection and segmentation, for the task. The network was trained on simulated bursts with single-component Gaussian profiles, broadband, and free from complex features, injected onto MeerKAT L-band noise observations. Pre-trained weights from the COCO dataset were used to initialise training, which was conducted over 40 epochs until overfitting became apparent.

Fixing the bow tie signal’s size and shape introduced the mask size as a tunable hyperparameter. We therefore trained models with varying mask sizes. Precision–recall analysis demonstrated that models with larger masks outperformed those with smaller masks. For the largest tested mask, the Average Precision (AP) was 0.997 at an Intersection over Union (IoU) threshold of 0.5, and 0.490 at IoU 0.95.

To evaluate performance on real data, we tested the model on bursts from nearby bright pulsars and FRBs detected across multiple MeerKAT frequency bands. All bursts were successfully identified, demonstrating the pipeline’s capability across varying DMs and observing configurations.

We further applied our model to MeerKAT follow-up observations of the repeater FRB20240114A, containing 62 bursts (18 in L-band and 44 in UHF). Our model detected 60 of these, including two bursts with S/N below the conventional threshold of eight—demonstrating the value of utilising more comprehensive information for transient searches. The two missed events were low-S/N bursts closely spaced in time to bright bursts, which limited detection due to the non-maximum suppression step.

Notably, many of the FRB bursts exhibited frequency structures not present in the training set—such as narrowband emission and deviations from a flat spectrum. Similarly, it also detected sequences of bursts visible in the same images. The fact that the model was able to identify them correctly highlights the robustness and adaptability of our approach.

\section*{Acknowledgements}
The authors would like to thank Raghuttam Hombal for his contributions to the development and maintenance of the GitHub repository. We also thank Jun Tian for providing a real dataset and sharing his findings, which enabled a direct comparison between the traditional and proposed pipelines.

The MeerKAT telescope is operated by the South African Radio Astronomy Observatory (SARAO), which is a facility of the National Research Foundation, an agency of the Department of Science and Innovation. We thank staff at SARAO for their help with observations and commissioning.

SBD acknowledges the support of a Science and Technology Facilities Council (STFC) stipend (grant number: ST/X001229/1) to permit work as a postgraduate researcher. 

BWS and RB were supported by a consolidated grant from STFC.

\section*{Data Availability}

Training and test data set is available upon request. The code is available at \url{https://github.com/sbelmontediaz/Hermes} .



\bibliographystyle{rasti}
\bibliography{bowties} 


\bsp	
\label{lastpage}
\end{document}